\documentclass[aps,prl,twocolumn,showpacs,superscriptaddress]{revtex4}
\usepackage{graphicx}

\begin{document}

\title{Low-lying quasiparticle states and hidden collective charge instabilities in parent cobaltate superconductors (Na$_x$CoO$_2$) }
\author{D. Qian} \affiliation{Department of Physics, Joseph
Henry Laboratories of Physics, Princeton University, Princeton, NJ
08544}
\author{D. Hsieh}
\author{L. Wray}
\affiliation{Department of Physics, Joseph Henry Laboratories of
Physics, Princeton University, Princeton, NJ 08544}
\author{A. Fedorov} \affiliation{Advanced Light Source, Lawrence Berkeley
Laboratory, Berkeley, CA 94305}
\author{D. Wu}
\author{J.L. Luo}
\author{N.L. Wang}
\affiliation{Institute of Physics, Chinese Academy of Sciences,
Beijing 100080, China}
\author{L. Viciu}
\author{R.J. Cava}
\affiliation{Department of Chemistry, Princeton University,
Princeton, NJ 08544}
\author{M.Z. Hasan}
\affiliation{Department of Physics, Joseph Henry Laboratories of
Physics, Princeton University, Princeton, NJ 08544}
\affiliation{Princeton Center for Complex Materials, Princeton
University, Princeton, NJ 08544}

\date{\today}

\begin{abstract}
We report a state-of-the-art photoemission (ARPES) study of high
quality single crystals of the Na$_x$CoO$_2$ series focusing on the
fine details of the low-energy states. The Fermi velocity is found
to be small ($<$ 0.5 eV.$\AA$) and only weakly anisotropic over the
Fermi surface at all dopings setting the size of the pair
wavefunction to be on the order of 10-20 nanometers. In the low
doping regime the exchange inter-layer splitting vanishes and two
dimensional collective instabilities such as 120$^{\circ}$-type
fluctuations become kinematically allowed. Our results suggest that
the unusually small Fermi velocity and the unique symmetry of
kinematic instabilities distinguish cobaltates from other
unconventional oxide superconductors such as the cuprates or the
ruthenates.

\end{abstract}


\maketitle

Research on strongly correlated electron systems has led to the
discovery of unconventional states of matter such as those realized
in the high T$_c$ superconductors, quantum Hall systems and
low-dimensional quantum magnets. Triangular cobaltates Na$_x$CoO$_2$
are a novel class of correlated electron systems with a rich phase
diagram. Superconductivity (near x=1/3) and correlated insulator
behavior (x=1/2) are observed in the low doping regime and enhanced
thermoelectric power (near x=2/3) and an unusual spin-density-wave
state (beyond x=3/4) are seen in the high doping regime
\cite{1}-\cite{3}. None of the existing theories of cobaltates can
account for the changes of electronic groundstates with doping, and
no systematic study of its \textit{low-energy} dynamics (e.g., Fermi
velocity, inter-layer coupling, particle-hole instabilities etc.)
exists so far.

An early ARPES study \cite{4} carried out on the host compound found
only one Fermi surface (FS), however the fine details of the
quasiparticles were not resolved. A subsequent study by Yang et.al.
\cite{5}, reported a surface state with a large FS and broad
bulk-representative quasiparticles due to the surface annealing
process but only one bulk FS was observed reconfirming the finding
in Ref.[4]. In this Letter, we report the fine details of the
low-energy quasiparticle states which allow us to determine the
dimensionality and the kinematic instabilities associated with the
electronic structure as a function of doping as well as the
fundamental parameters to compare the electron dynamics with other
unconventional superconductors. Such a study is made possible due to
the lack of surface reconstruction (hence no surface state
complications in the interpretation of the data) in our high quality
crystals. The strong interlayer coupling we resolve in the high
doping regime provides evidence for a natural mechanism for the
unusual magnetic order observed, even though the system is far away
from the conventional half-filled Mott limit (x=0). In contrast,
fine details of the low-energy behavior in the low doping regime (x
$=$ 1/3) reveal a hidden two dimensional particle-hole instability
predicted in recent many-body theories. Moreover, the results allow
us to reliably compare the low-energy parameters obtained on
cobaltates with other materials classes and provide fundamental
ingredients for developing a microscopic theory of these materials.

\begin{figure}[t]
\center \includegraphics[width=8cm]{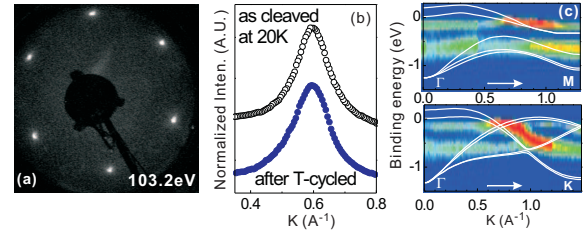} \caption{(a) Low
energy electron diffraction image of the cleaved (001) surface
exhibits hexagonal symmetry of the cobalt layers. No Ruthenate-like
surface reconstruction is observed. (b) Temperature cycling (20K
$\leftrightarrow$ 100K) was found to have no effect on the momentum
distribution (MDCs) of the quasiparticles. Only one peak is observed
at x=0.57 and lower dopings. (c) The valence band overlayed with
band calculations \cite{6} shows narrowing due to electron
correlations.}
\end{figure}

\begin{figure*}[t]
\center \includegraphics[width=14cm]{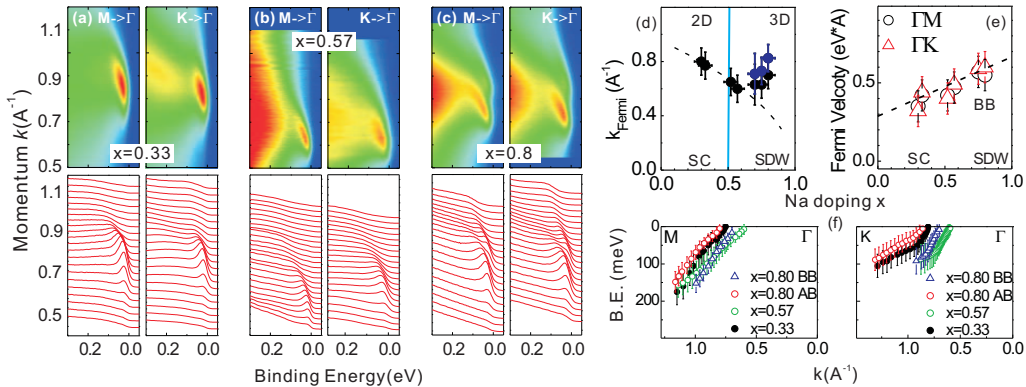}
\caption{\textbf{Evolution of low-lying quasiparticle states}:
Single-particle removal spectra for x=0.33 (a), x=0.57 (b) and
x=0.80 (c) along the $\Gamma$-M and $\Gamma$-K cuts. Low doped
samples (a,b) exhibit only one Fermi crossing while high doped
samples (c) show two Fermi crossings. Doping evolution of Fermi
surface size ($k_{F}$) is shown in (d) and Fermi velocity in (e).
Fermi velocity increases with increasing doping. Bonding (BB) and
antibonding (AB) bands are identified at high doping. Note that the
average (BB+AB) Fermi velocity \textit{decreases} in approaching
x=1. The dotted line in (d) is calculated based on the 2-D Luttinger
theorem \cite{4}. (f) Dispersion plots along the M-$\Gamma$ and
K-$\Gamma$ cuts.}
\end{figure*}

Spectroscopic measurements were performed with 30 eV to 90 eV
photons with better than 10 to 25 meV energy resolution, and angular
resolution better than 0.8\% of the Brillouin zone at ALS Beamlines
12.0.1 and 10.0.1, using Scienta analyzers with chamber pressures
below 4$\times$10$^{-11}$ torr. High quality single crystals over a
wide doping range x= 0.3, 0.33, 0.52, 0.57(Na/K), 0.7, 0.75 and 0.79
were used for this study. Cleaving the samples in situ at 20 K (or
100K) resulted in shiny flat surfaces, characterized by diffraction
to be clean and well ordered with the same symmetry as the bulk
(Fig-1). No surface state was observed. All presented data were
taken at 20K, although in a few cases samples were studied at 100K
for cross-checking.

Fig-2 shows the doping dependence of single electron removal spectra
as a function of energy and momentum. For low doping (x $<$ 2/3),
one quasiparticle feature is seen to disperse from high binding
energies at high momentum values near the corner (K) or the face (M)
of the reciprocal space to the Fermi level. High resolution study of
the quasiparticle feature shows that it is well separated from the
hump structure observed at 200-300 meV binding energies at all
doping. The quasiparticle lifetime drops very fast making its
intensity vanish beyond 70 to 100 meV binding energies depending on
the doping. The low energy (0-50 meV) Fermi velocities extracted
from the data are shown in Fig-2(e). The Fermi velocity is found to
be weakly anisotropic, (e.g., similar in magnitude along $\Gamma$-M,
and $\Gamma$-K directions) at all doping levels, and increases with
doping away from the Mott limit (x=0). The Fermi velocity, averaged
over the FS, is about 0.37 eV$\cdot$\AA  for x =0.3. Given the size
of the FS, we estimate the carrier mass, m* $\sim$
$\hbar$k$_F$/$|v_F|$ $\sim$ 15 to 30 m$_e$. This is rather large
compared to most known transition metal oxide superconductors.
However, a similarly large carrier mass (2m$^*$ $\leq$ 70 m$_e$) has
recently been reported by $\mu$SR measurements\cite{7}. Fig-2(d)
plots the average size of the Fermi surface (k$_F$) as a function of
doping. Doping evolution is found to follow  the 2-D Luttinger
theorem (FS area $\propto$ (1-x)) up to x near 2/3. The Fermi
surface (k$_F$) gets broadened and splits into two clearly separate
ones beyond x=0.7.

Fig-3(a) shows the momentum-distribution (n(k)-plot) for x=0.33 as a
representative of the low-doping regime. The inner edge of this
density plot is the Fermi surface. The electron distribution in the
extended zone scheme is shown in Fig-3(b). In Fig-4 we show the
quasiparticle behavior in a typical high doping (x $>$ 2/3) sample.
Two quasiparticles are observed to cross the Fermi level with fine
k-splitting. A systematic study shows coupled oscillatory behavior
of quasiparticle peak intensities with increasing incident photon
energy. This is commonly observed for states with nearly orthogonal
symmetries \cite{8}. The unit cell of Na$_x$CoO$_2$ has two
Co-layers - therefore the Co-derived bands with orthogonal
symmetries are expected to be split, leading to two Fermi crossings
similar to what is observed in multi-layer materials \cite{8,9}.
Inter-layer splitting in cobaltates has been predicted by LDA band
theory \cite{6}. It is known from neutron and x-ray diffraction
studies that with increasing Na content the separation between the
two cobalt layers decreases substantially \cite{10,11}. Filling of
the Na layer by high Na density (high doping) leads to stronger
bonding between the Co layers resulting in decreased interplanar
(c-axis) distance. Therefore, our observation of inter-layer
splitting at high doping and its absence at low doping suggest a two
to three dimensional crossover of the low-lying electronic structure
with increasing sodium concentration. The large interlayer splitting
observed only at high x suggests that c-axis exchange is large,
since interlayer hopping and magnetic exchange (exchange
inter-layer) are directly related (J$_{\bot}$ $\propto$
t$_{\bot}$$^2$). The crucial role of the Na-layer (atomic
rearrangements) at high doping in determining the magnetic state of
the sample is thus consistent with recent NMR findings \cite{13}.

The Fermi surface area at high doping does not exactly match the 2-D
Luttinger count (deviation of data from the dotted line in
Fig-2(d)). The fact that the Fermi velocity and other band behavior
exhibit systematic changes with doping suggests that the deviation
can not be due to doping inhomogeneity. We attribute this deviation
in FS area to two factors which were overlooked in earlier studies.
The SDW-order leads to a canonical doubling of the primitive unit
cell, and since the order is fully 3-D, a projected or quasi-2D map
of the FS does not necessarily capture the full volume. Thus the 2-D
Luttinger count is not applicable to the FS of highly doped
cobaltates, where a 3-D SDW is observed. This fact and the
unresolved bilayer splitting provide clues to the puzzle of the
reported FS being larger in previous studies \cite{4,5}. To perform
a proper Luttinger count one would need to count the full 3D Fermi
surface \textit{volume} and consider the effect of the full SDW
order on the electronic structure.

\begin{figure}[t]
\center \includegraphics[width=9cm]{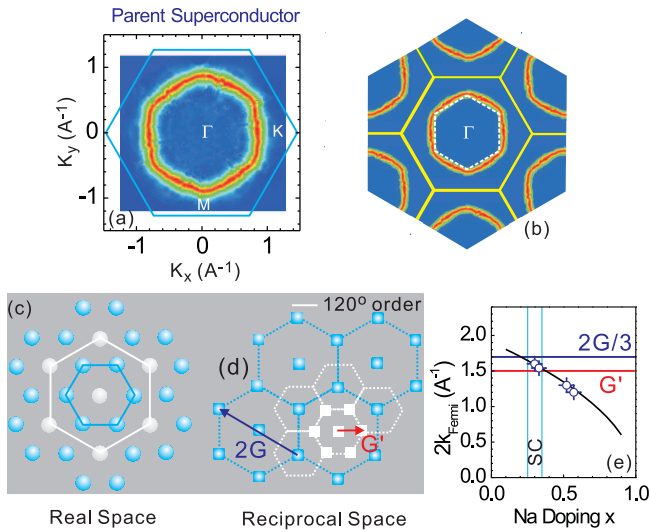}
\caption{\textbf{Fermi Surface and Collective Instabilities}: (a)
Momentum distribution of electrons in the x=0.33 sample measured in
a simultaneous azimuthal scan mode with an energy window of 10 meV.
(b) Electron distribution shown in an extended zone scheme. (c,d)
Real-space arrangement of cobalt atoms and 120$^{\circ}$ type order
and their corresponding reciprocal lattices. (e) Average
electron-hole excitation wave-vector (2k$_F$) as a function of Na
doping x. In the low doping regime, the measured data points (open
circles) fall on the 2-D Luttinger theorem (solid) line. Horizontal
solid lines represent two commensurate instabilities. The
120$^{\circ}$ order line ($G'$, red) intersects the (2k$_F$)-line
near x=0.34 and the 2G/3 lattice vector line intersects near x=0.25
enclosing the superconducting phase boundaries \cite{22}.}
\end{figure}

\begin{figure}[h]
\center \includegraphics[width=8cm]{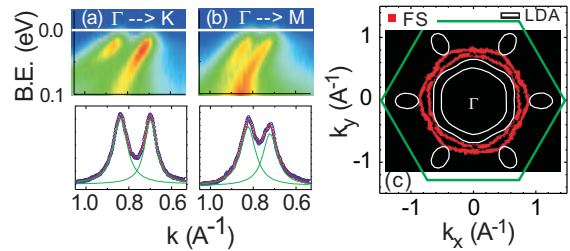} \caption{(a,b) A
two-band (double) crossing behavior is observed for x$>$ 2/3.
Systematics of data are shown for x=3/4 where samples exhibit 3-D
SDW order. The data can be modeled with two Lorentzians. (c) The
Fermi surface topology is compared with inter-layer splitting in LDA
calculation\cite{6}. The observed FS "consumes" the area of the
corner pockets.}
\end{figure}

\begin{table*}[t]
\caption{Comparison of cobaltates with Bardeen-Cooper-Shrieffer
(BCS) and non-BCS cuprate superconductors}%
\begin{ruledtabular}
\begin{tabular}{|c|c|c|c|c|c|c|c|}
&&&Retardation[17]&Phase&Mass
 &Size of the&\\
Class&T$_c$(K)&${Fermi Velocity}$&$\frac{E_{fermi}(W)}{\hbar\omega_{Ph(optic)}}$&Ordering[17,18,21]&$\frac{\hbar k_{fermi}}{\nu_{fermi}}$&Pair Wfn ($\AA$)& Ref.\\
&&$(eV.{\AA})$&(by Phonons)&T$_\theta (K) \sim
n_{2d(SF)}/m^*$&m$^*$/m$_e$&$\sim
0.2\frac{\hbar\nu_{fermi}}{k_BT_c}$&\\
 \hline
Cobaltates (NaCoO)
&5&\textbf{0.37$\pm$0.1}&\textbf{$<$4}&\textbf{$<$20}&
\textbf{$>$15}&$\sim$200&\textbf{Present work}\\
\hline p-Cuprates (LSCO)&38&1.8&8&54&2(nodal)&$\sim$100&[16,17] \\
\hline n-Cuprates (NCCO)&21&2.0&9&130&2.4(nodal)&$\sim$210&[16,17]\\
\hline MgB$_2$ &39&2-7&10$^2$&10$^3$&x&x&[19]\\
\hline Lead (Pb)&7.2&10.4&10$^3$&10$^5$&1.8&10$^4$&[17,20]\\
\end{tabular}
\end{ruledtabular}
\end{table*}

Based on the vanishing of interlayer coupling in going from high to
low doping, we conclude that the electronic structure is largely two
dimensional near the superconducting Na concentration. We now
discuss the possibility of electron-hole excitation induced
instabilities with 2-D commensurate wave vectors. We note that the
2-D FS near x=1/3 exhibits nearly straight sections perpendicular to
the high symmetry directions. At low doping, this distance 2k$_F$ is
very close to 2$\bf{G}$/3, where $\bf{G}$ is the fundamental
reciprocal lattice vector (Fig-3). However, this vector can not nest
the pieces of the FS we observe. Another instability is that of the
120$^{\circ}$ type which can arise from weak spatial modulations of
charge, spin or orbital densities. Our data shows that the
electronic system in the vicinity of x=1/3 is susceptible to such
kinematic instabilities. This is due to the fact that the BZ of such
order or fluctuations coincides with the topology of our measured
FS. The geometrical construct for 120$^{\circ}$ order is shown in
real and reciprocal space in Fig-3(c) and 3(d). The Brillouin zone
constructed out of the white hexagons in Fig-3(d) is
\textit{superimposed} on the measured Fermi surface (white hexagon)
presented in Fig-3(b). The coincidence, including the k-space
anisotropies, can be noted in Fig-3(b). For this type of instability
there is no specific single nesting vector and the instability is
\textit{collective} in nature involving \textit{all} parts of the
FS. In order to determine whether or not this instability is unique
to x=0.3, we have carried out the doping dependence of FS size in
the vicinity of x=0.3. The uniqueness of this instability ($G'$) can
be seen in Fig-3(e) - the $G'$= 2$k_{F}$ line intersects the
Luttinger line uniquely near x=0.34, which is the doping for
superconductivity \cite{22}.

The doping evolution of electron behavior and low-energy correlation
effects have been theoretically studied in an extended Hubbard model
on a triangular lattice\cite{14}. These studies predict strong
renormalization of bandwidth (i.e., small Fermi velocity) over the
phase diagram. Moreover, a strictly two dimensional
120$^{\circ}$-type collective instability or order, such as
$\sqrt{3}\times\sqrt{3}$, near x =1/3 and 2/3 is predicted. Our
results show that only near x=1/3 is such a \textit{two dimensional}
instability (with triangular symmetry) allowed with the measured
topology of the Fermi surface (Fig-3). Near x=2/3 on the other hand,
electronic structure is \textit{three dimensional}. Theory suggests
that if this order is long-range, the quasiparticle spectrum would
be gapped at all momenta (k) and the Fermi velocity would exhibit
strong renormalization at nearby dopings. In our study, no gap
opening is observed in the quasiparticle dispersion (Fig-2(a)) at
any momentum at this doping down to the base temperature of the
experiment. However, the Fermi velocity is seen to be suppressed
over a range of doping (Fig-2(e)). Therefore, our results suggest a
strongly \textit{fluctuating} character of this collective
instability \cite{22}.

The strong correlation models have further been considered in
connection with superconductivity \cite{14,15}. These studies
suggest that a kinematic or \textit{hidden} FS instability with
strong fluctuations can lead to unconventional superconductivity
with an \textit{f}-wave order parameter, even within a Fermi-liquid
quasiparticle-like picture. The quasiparticle parameters related to
the cobalt derived states with triangular symmetry are therefore of
interest for theory. Based on our data, we estimate the parameters
in cobaltates presented in Table-I. For comparison we also quote
similar parameters obtained on conventional materials as well as the
unusual materials classes. For use in Table-I, the Fermi energy of
cobaltates is estimated to be on the order of the occupied bandwidth
(following the approximation in ref.\cite{17}) which is taken from
our data in Fig-2(f). At the order-of-magnitude level, our results
show that the cobaltate exhibits a lack of a retardation effect due
to its small Fermi energy, possesses a small phase ordering (kinetic
energy) scale due to its large effective mass, and exhibits a
relatively short coherence length (Cooper pair wavefunction) due to
its small Fermi velocity (consequence of uncertainty relation). All
these characteristics can be traced to the strong renormalization of
the Fermi velocity which, in cobaltates, is about a factor of five
smaller than it is in the high T$_c$ cuprates. Given such character
for the low-lying states and the k-space kinematics we observe, it
is unlikely that a conventional mechanism is at play.

In conclusion, fine details of the low-energy states in
Na$_x$CoO$_2$ are resolved due to high crystalline quality of
materials. A splitting of the single-particle band signalling
inter-layer coupling is observed in the high doping regime which
accounts for the observed three dimensional magnetism. In the low
doping regime, in contrast, the splitting disappears leading to the
two dimensional character of triangularly correlated electron
motion. Our results suggest that the strongly renormalized Fermi
velocity and the unique two dimensional symmetry of kinematic
instabilities distinguish cobaltates from most other oxide
superconductors and clearly so from BCS superconductors. The
quantitative details of the low-lying states in our studies
(Table-I) provide important guides to develop a comprehensive theory
of cobaltates.

We gratefully acknowledge P.W. Anderson, D.A. Huse, S.A. Kivelson,
P.A. Lee, N. P. Ong, S. Shastry and S. Sondhi for discussions. This
work is partially supported through the NSF(DMR-0213706), DOE(
DE-FG02-05ER46200 and DE-FG02-98-ER45706) and NSFC(10574158).


\begin{thebibliography}{99}
\bibitem{21} Since the superfluid density is not known we have used
the Fermi sea to calculate the phase ordering scale at the order of
magnitude level.
\bibitem{22} We note a natural enlargement of the central FS at x=1/3 in our data (Fig-3) with respect to
LDA predictions given the absence of corner FS pockets (Ref.[4,5])
and the constraint of the Luttinger theorem. Therefore,
120$^{\circ}$ type kinematic instabilities are seen to be only
allowed in the absence of the corner pockets. Remarkably, our
results suggest that both the suppression of corner pockets and the
enhanced susceptibility to 120$^{\circ}$ type order are natural
consequences of long-range Coulomb interactions.
\end{thebibliography}
\end{document}